\documentclass[11pt]{article}

\usepackage{amsfonts}
\usepackage{amsmath}
\usepackage{amssymb}
\usepackage{amsthm}

\usepackage{graphicx}

\usepackage{caption2}

\makeatletter \oddsidemargin-.25in \evensidemargin-.25in
\makeatother

\usepackage{graphicx}

\theoremstyle{plain}

\theoremstyle{definition}

\begin{document}
\title{Origin and evolution of the genetic code: The universal enigma}

\author{Eugene V. Koonin\footnote{e-mail: koonin@ncbi.nlm.nih.gov}\, and Artem S. Novozhilov\\[2mm]
{\small \textit{National Center for Biotechnology Information,}}\\ {\small \textit{National Library of Medicine, National Institutes of Health, Bethesda, MD 20894}}} 

\date{}

\maketitle

\begin{abstract}
The genetic code is nearly universal, and the arrangement of the codons in the standard codon table is highly non-random. The three main concepts on origin and evolution of the code are the stereochemical theory, according to which codon assignments are dictated by physico-chemical affinity between amino acids and the cognate codons (anticodons); the coevolution theory, which posits that the code structure coevolved with amino acid biosynthesis pathways; and the error minimization theory under which selection to minimize the adverse effect of point mutations and translation errors was the principal factor of the code's evolution.  These theories are not mutually exclusive and are also compatible with the frozen accident hypothesis, i.e., the notion that the standard code might have no special properties but was fixed simply because all extant life forms share a common ancestor and remained, mostly, unchanged because of the deleterious effect of codon reassignment. Mathematical analysis of the structure and possible evolutionary trajectories of the code shows that it is highly robust to translational misreading but there are a huge number of more robust codes, so that the standard code potentially could evolve from a random code via a short sequence of codon series reassignments. Thus, much of the evolution that led to the standard code can be interpreted as a combination of frozen accident with selection for error minimization although contributions from coevolution of the code with metabolic pathways and/or weak affinities between amino acids and nucleotide triplets cannot be ruled out. However, such scenarios for the code evolution are based on formal schemes whose relevance to the actual primordial evolution is uncertain, so much caution in interpretation is necessary. A real understanding of the code's origin and evolution is likely to be attainable only in conjunction with a credible scenario for the evolution of the coding principle itself and the translation system.

\end{abstract}


\section{Introduction}
\begin{figure}[bth]
\centering
\includegraphics[width=0.6\textwidth]{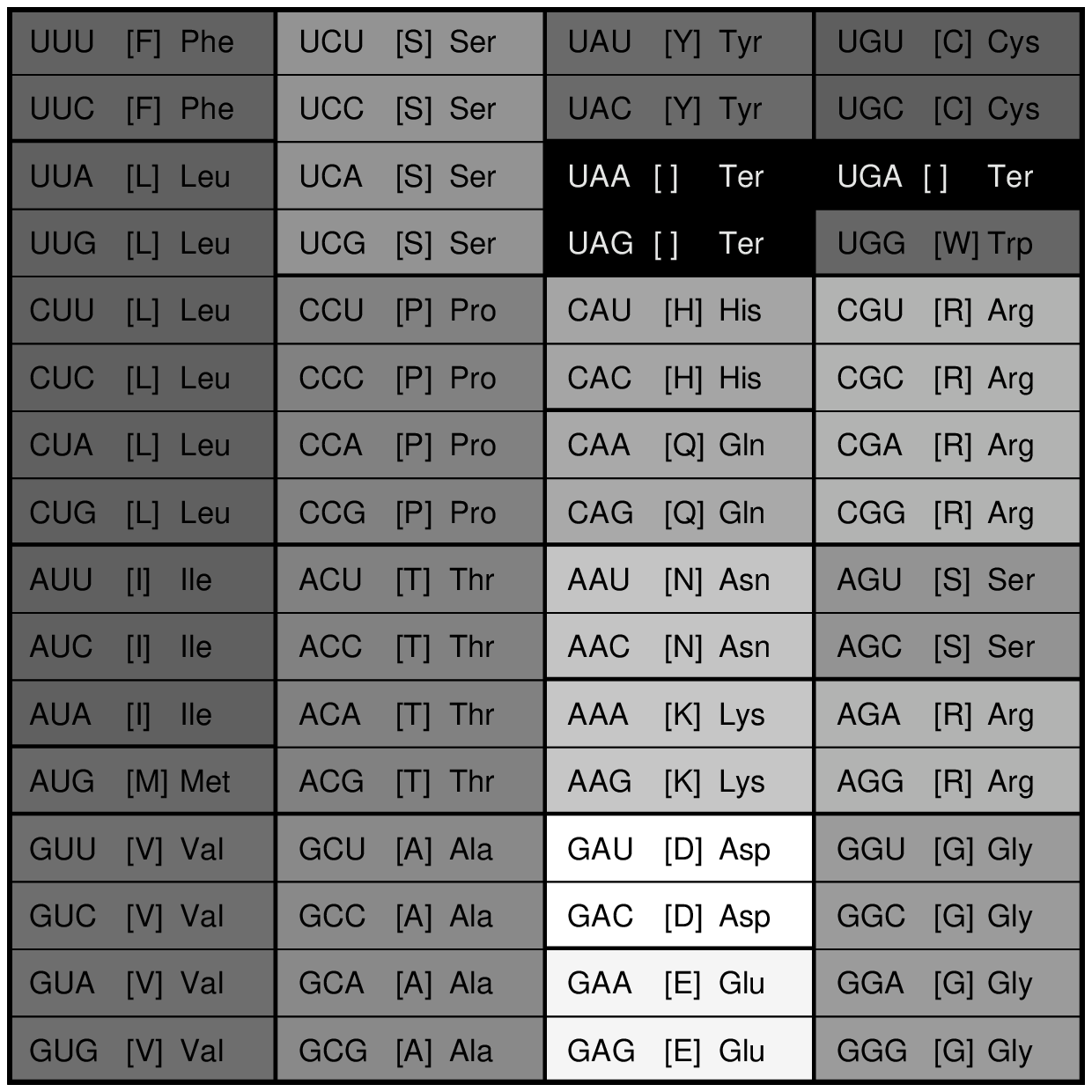}
\caption{The standard genetic code. The codon series are shaded in accordance with the polar requirement scale values (Woese et al. 1966b), which is a measure of an amino acid's hydrophobicity: the greater hydrophobicity the darker the shading (the stop codons are shaded black).}
\end{figure}

Shortly after the genetic code of \textit{Escherichia coli} was deciphered (Nirenberg et al. 1963), it was recognized that this particular mapping of 64 codons to 20 amino acids and two punctuation marks (start and stop signals) is shared, with relatively minor modifications, by all known life forms on earth (Hinegardner and Engelberg 1963; Woese, Hinegardner, and Engelberg 1964). Even a perfunctory inspection of the standard genetic code table (Fig. 1) shows that the arrangement of amino acid assignments is manifestly nonrandom (Woese 1965a; Woese 1967; Crick 1968; Ycas 1969). Generally, related codons (i.e., the codons that differ by only one nucleotide) tend to code for either the same or two related amino acids, i.e., amino acids that are physico-chemically similar (although there are no unambiguous criteria to define physicochemical similarity). The fundamental question is how these regularities of the standard code came into being, considering that there are more than $10^{84}$ possible alternative code tables if each of the 20 amino acids and the stop signal are to be assigned to at least one codon. More specifically, the question is, what kind of interplay of chemical constraints, historical accidents, and evolutionary forces could have produced the standard amino acid assignment, which displays many remarkable properties. The features of the code that seem to require a special explanation include, but are not limited to, the block structure of the code, which is thought to be a necessary condition for the code's robustness with respect to point mutations, translational misreading, and translational frame shifts (Chechetkin 2003); the link between the second codon letter and the properties of the encoded amino acid so that codons with U in the second position correspond to hydrophobic amino acids (Rumer 1966; Vol'kenshtein and Rumer 1967); the relationship between the second codon position and the class of aminoacyl-tRNA synthetase (Wetzel 1995), the negative correlation between the molecular weight of an amino acid and the number of codons allocated to it (Hasegawa and Miyata 1980; Di Giulio 2005); the positive correlation between the number of synonymous codons for an amino acid and the frequency of the amino acid in proteins (King and Jukes 1969; Gilis et al. 2001); the apparent minimization of the likelihood of mistranslation and point mutations (Haig and Hurst 1991; Freeland, Wu, and Keulmann 2003); and the near optimality for allowing additional information within protein coding sequences (Itzkovitz and Alon 2007).

When considering the evolution of the genetic code, we proceed under several basic assumptions that are worth spelling out. It is assumed that there are only 4 nucleotides and 20 encoded amino acids (with the notable exception of selenocysteine and pyrrolysine, for which subsets of organisms have evolved special coding schemes (Ambrogelly, Palioura, and Soll 2007), see also discussion below) and that each codon is a triplet of nucleotides. It has been argued that movement in increments of three nucleotides is a fundamental physical property of RNA translocation in the ribosome so that the translation system originated as a triplet-based machine (Aldana et al. 1998; Aldana-Gonzalez et al. 2003; Gusev and Schulze-Makuch 2004). Obviously, this does not rule out the possibility that, e.g., only two nucleotides in each codon are informative (see, e.g., (Patel 2005; Wu, Bagby, and van den Elsen 2005; Travers 2006; Ikehara and Niihara 2007) for hypotheses on the evolution of the code through a ``doublet'' phase). Questions on why there are four standard nucleotides in the code (Szathmary 1991; Szathmary 2003) or why the standard code encodes 20 amino acids (Weber and Miller 1981; Lu and Freeland 2006; Lu and Freeland 2008) are fully legitimate. Conceivably, theories on the early phases of the evolution of the code should be constrained by the minimal complexity that is required of a self-replicating system (e.g., (Munteanu et al. 2007)). However, this fascinating are of enquiry is beyond the scope of this review, and for the present discussion we adopt the above fundamental numbers as assumptions. With these premises, we here attempt to critically assess and synthesize the main lines of evidence and thinking about the code's nature and evolution.

\section{The code is evolvable}
The code expansion theory proposed in Crick's seminal paper posits that the actual allocation of amino acids to codons is mainly accidental and ``yet related amino acids would be expected to have related codons'' (Crick 1968). This concept is known as ``frozen accident theory'' because Crick maintained, following the earlier argument of Hinegardner and Engelberg (Hinegardner and Engelberg 1963) that, after the primordial genetic code expanded to incorporate all 20 modern amino acids, any change in the code would result in multiple, simultaneous changes in protein sequences and, consequently, would be lethal, hence the universality of the code. Today, there is ample evidence that the standard code is not literally universal but is prone to significant modifications, albeit without change to its basic organization.

Since the discovery of codon reassignment in human mitochondrial genes (Barrell, Bankier, and Drouin 1979), a variety of other deviations from the standard genetic code in bacteria, archaea, eukaryotic nuclear genomes and, especially, organellar genomes have been reported, with the latest census counting over 20 alternative codes (Knight, Freeland, and Landweber 1999; Knight, Freeland, and Landweber 2001; Yokobori, Suzuki, and Watanabe 2001; Santos et al. 2004; Sengupta, Yang, and Higgs 2007). All alternative codes are believed to be derived from the standard code (Knight, Freeland, and Landweber 2001); together with the observation that many of the same codons are reassigned (compared to the standard code) in independent lineages (e.g., the most frequent change is the reassignment of the stop codon UGA to tryptophan), this conclusion implies that there should be predisposition towards certain changes; at least one of these changes was reported to confer selective advantage (Santos et al. 1999).

The underlying mechanisms of codon reassignment typically include mutations in tRNA genes, where a single nucleotide substitution directly affects decoding (Giege, Sissler, and Florentz), base modification (Matsuyama et al. 1998), or RNA editing (Alfonzo et al. 1999) (reviewed in (Knight, Freeland, and Landweber 2001)). Another pathway of code evolution is recruitment of non-standard amino acids. The discovery of the $21^\textrm{st}$ amino acid, selenocysteine, and the intricate molecular machinery that is involved in the incorporation of selenocysteine into proteins (Allmang and Krol 2006) initially has been considered a proof that the current repertoire of amino acids is extremely hard to change. However, the subsequent discovery of the second non-canonical amino acid, pyrrolysine, and, importantly, the existence of a pyrrolysine-specific tRNA revealed additional malleability of the code (Krzycki 2005; Ambrogelly, Palioura, and Soll 2007). In addition to the variations on the standard code discovered in organisms with minimized genomes, many experimental attempts on code modification and expansion have been reported (Wang, Xie, and Schultz 2006). Recently, a general method has been developed to encode the incorporation of unnatural amino acids in genomes by recruiting either one of the stop codons or a subset of a codon series for a particular amino acid and engineering the cognate tRNA and aminoacyl-tRNA synthetase (Xie and Schultz 2006). The application of this methodology has already allowed incorporation in \textit{E. coli} proteins of over 30 unnatural amino acids, in a striking demonstration of the potential malleability of the code (Wang, Xie, and Schultz 2006; Xie and Schultz 2006).

Three major theories have been suggested to explain the changes in the code. The ``codon capture'' theory (Osawa et al. 1992; Osawa 1995) proposes that, under mutational pressure to decrease genomic GC-content, some GC-rich codons might disappear from the genome (particularly, a small, e.g., organellar, genome). Then, due to random genetic drift, these codons would reappear and would be reassigned as a result of mutations in non-cognate tRNAs. This mechanism is essentially neutral, i.e., codon reassignment would occur without generation of aberrant or non-functional proteins.
 
Another concept of code alteration is the ``ambiguous intermediate'' theory which posits that codon reassignment occurs through an intermediate stage where a particular codon is ambiguously decoded by both the cognate tRNA and a mutant tRNA (Schultz and Yarus 1994; Schultz and Yarus 1996). An outcome of such ambiguous decoding and the competition between the two tRNAs could be eventual elimination of the gene coding for the cognate tRNA and takeover of the codon by the mutant tRNA (Santos et al. 2004; Chechetkin 2006). The same mechanism might also apply to reassignment of a stop codon to a sense codon, when a tRNA that recognizes a stop codon arises by mutation and captures the stop codon from the cognate release factor. Under the ambiguous intermediate hypothesis, a significant negative impact on the survival of the organism could be expected but the finding that the CUG codon (normally coding for leucine) in the fungus \textit{Candida zeylanoides} is decoded as either leucine (3-5\%) or serine (95-97\%) gave credence to this scenario (Suzuki, Ueda, and Watanabe 1997; Santos et al. 2004).
 
Finally, evolutionary modifications of the code have been linked to ``genome streamlining'' (Andersson and Kurland 1995; Andersson and Kurland 1998). Under this hypothesis, the selective pressure to minimize mitochondrial genomes yields reassignments of specific codons, in particular, one of the three stop codons.
 
The three theories explaining codon reassignment are not exclusive considering that the ``ambiguous intermediate'' stage can be preceded by a significant decrease in the content of GC-rich codons, so that codon reassignment might be driven by a combination of evolutionary mechanisms (Massey et al. 2003), often under the pressure for genome minimization, especially, in organellar genomes and small genomes of parasitic bacteria such as mycoplasmas (Andersson and Kurland 1991; Andersson and Kurland 1998; Massey and Garey 2007; Sengupta, Yang, and Higgs 2007).

\section{The basic theories of the code nature, origin and evolution}

The existence of variant codes and the success of experiments on the incorporation of unnatural amino acids briefly discussed in the preceding section indicates that the genetic code has a degree of evolvability. However, all these deviations involve only a few codons, so in its main features, the structure of the code seems not to have changed through the entire history of life or, more precisely, at least, since the time of the Last Universal Common Ancestor (LUCA) of all modern (cellular) life forms. This universality of the genetic code and the manifest non-randomness of its structure cry for an explanation(s). Of course, Crick's frozen accident/code expansion theory can be considered a default explanation that does not require any special mechanisms and is only predicated on the existence of a LUCA with an advanced translation system resembling the modern one (that is, the implicit assumption is that LUCA was not a ``progenote'' with primitive, very inaccurate translation (Woese and Fox 1977)). However, this explanation is often considered unsatisfactory, first, on the most general, epistemological grounds, because it is, in a sense, a non-explanation, and second, because the existence of variant codes and the additional, experimentally revealed flexibility of the code (see above) present a challenge to the frozen-accident view. Indeed, the fact that there seem to be ways to ``sneak in'' changes to the standard code, and yet, the same limited modifications seem to have evolved independently in diverse lineages, suggests that the code structure could be non-accidental. Three, not necessarily mutually exclusive main theories have been proposed in attempts to attribute the pattern of amino acid assignments in the standard genetic code to physico-chemical or biological factors or a combination thereof. Rather remarkably, the central ideas of each of these theories have been formulated during the classic age of molecular biology, not long after the code was deciphered or even earlier, and despite numerous subsequent developments, remain relevant to this day. We first briefly outline the three theories in their respective historical contexts and then discuss the current status of each.

\begin{enumerate}

\item The \textit{stereochemical theory} asserts that the codon assignments for particular amino acids are determined by a physicochemical affinity that exists between the amino acids and the cognate nucleotide triplets (codons or anticodons). Thus, under this class of models, the specific structure of the code is not at all accidental but, rather, necessary and, possibly, unique. The first stereochemical model was developed by Gamow in 1954, almost immediately after the structure of DNA has been resolved and, effectively, along with the idea of the code itself (Gamow 1954). Gamow proposed an explicit mechanism to relate amino acids and rhomb-shaped ``holes'' formed by various nucleotides in DNA. Subsequently, after the code was deciphered, more realistic stereochemical models have been proposed (Pelc 1965; Dunnill 1966; Pelc and Welton 1966) but were generally deemed improbable due to the failure of direct experiments to identify specific interactions between amino acids and cognate triplets (Woese 1967; Crick 1968). Nevertheless, the inherent attractiveness of the stereochemical theory, which, if valid, makes it much easier to see how the code evolution started, stimulated further experimental and theoretical activity in this area. 

\item The \textit{adaptive theory} of the code evolution postulates that the structure of the genetic code was shaped under selective forces that made the code maximally robust, i.e., minimize the effect of errors on the structure and function of the synthesized proteins. It is possible to distinguish the ``lethal-mutation'' hypothesis (Sonneborn 1965; Epstein 1966) under which the standard code evolved to minimize the effect of point mutations and the ``translation-error minimization'' hypothesis (Woese 1965b; Goldberg and Wittes 1966), which posits that the most important pressure in the code's evolution was selection for minimization of the effect of the translational misreadings.

A combination of the two types of forces is conceivable as well. The fact that related codons code for similar amino acids and the experimental observations that mistranslation occurs more frequently in the first and third positions of codons whereas it is the second position that correlates best with amino acid properties were construed as evidence in support of the adaptive theory (Davies, Gilbert, and Gorini 1964; Friedman and Weinstein 1964; Woese 1965b). The translation-error minimization hypothesis also received some statistical support from Monte Carlo simulations (Alff-Steinberger 1969), which later became a major tool to analyze the degree of optimization of the standard code.
\item The \textit{coevolution theory} posits that the structure of the standard code reflects the pathways of amino acid biosynthesis (Wong 1975). According to this scenario, the code coevolved with the amino acid biosynthetic pathways, i.e., during the code evolution, subsets of codons for precursor amino acids have been reassigned to encode product amino acids. Although the basic idea of the coevolution hypothesis is the same as in Crick's scenario of code extension, the explicit identification of precursor-product pairs of amino acids and strong statistical support for the inferred precursor-product pairs (Wong 1975; Wong 2005) gained the coevolution theory wide acceptance. 
\end{enumerate}

A complementary approach to the problem of code evolution espouses a ``tRNA-centric'' view under which the features of the code are determined by different types of co-evolution, namely, that of the codons and the cognate tRNA anticodons (Chechetkin 2006) or of the codons and aminoacyl-tRNA synthetases (Delarue 2007). This coevolution has been interpreted, primarily, in terms of minimization of the rate and effect of translation errors (Chechetkin 2006) or with respect to the reduction of coding ambiguity at the early stages of the code evolution (Delarue 2007).

\section{The stereochemical theory: tantalizing hints but no conclusive evidence}

Extensive early experimentation has detected, at best, weak and relatively non-specific interactions between amino acids and their cognate triplets (Woese et al. 1966a; Woese 1967; Saxinger, Ponnamperuma, and Woese 1971). Nevertheless, it is not unreasonable to argue that even a relatively weak, moderately selective affinity between codons (anticodons) and the cognate amino acids could have been sufficient to precipitate the emergence of the primordial code that subsequently evolved into the modern code in which the specificity is maintained by much more precise and elaborate, indirect mechanisms involving tRNAs and aminoacyl-tRNA synthetases. Furthermore, it can be argued that interaction between amino acids and triplets is strong enough for detection only within the context of specific RNA structures that ensure the proper conformation of the triplet; this could be the cause of the failure of straightforward experiments with trinucleotides or the corresponding polynucleotides. Indeed, the modern version of the stereochemical theory, the ``escaped triplet theory'' posits that the primordial code functioned through interactions between amino acids and cognate triplets that resided within amino-acid-binding RNA molecules (Yarus, Caporaso, and Knight 2005). The experimental observations underlying this theory are that short RNA molecules (aptamers) selected from random sequence mixtures by amino-acid-binding were significantly enriched with cognate triplets for the respective amino acids (Knight and Landweber 1998; Knight, Landweber, and Yarus 2003). Among the 8 tested amino acids (phenylalanine, isoleucine, histidine, leucine, glutamine, arginine, tryptophan, and tyrosin) only glutamine showed no correlation between the codon and the selected aptamers(Yarus, Caporaso, and Knight 2005). The straightforward statistical test applied in these analyses indicated that the probability to obtain the observed correlation between the codons and the sequences of the selected aptamers due to chance was extremely low; the most convincing results were seen for arginine (Yarus, Caporaso, and Knight 2005). However, more conservative statistical procedures (applied to earlier aptamer data) suggest that the aptamer-codon correlation could be a statistical artifact (Ellington, Khrapov, and Shaw 2000) (but see (Knight and Landweber 2000)).

A different kind of statistical analysis has been employed to calculate how unusual is the standard code, given the aptamer-amino-acid binding data (Knight, Landweber, and Yarus 2003; Yarus, Caporaso, and Knight 2005). A comparison of the standard code with random alternatives has shown that only a tiny fraction of random codes displayed a stronger correlation with the aptamer selection data than the standard code (the real genetic code has greater codon association than 90.3\% random codes, and greater anticodon association than 99.8\% random codes). The premises of this calculation can be disputed, however, because the standard code has a highly non-random structure, and one could argue that only comparison with codes of similar structures are relevant, in which case the results of aptamer selection might not come out as being significant. 

On the whole, it appears that the aptamer experiments, although suggestive, fail to clinch the case for the stereochemical theory of the code. As noticed above, the affinities are rather weak, so that even the conclusions on their reality hinge on the adopted statistical models. Even more disturbing, for different amino acids, the aptamers show enrichment for either codon or anticodon sequence or even for both (Yarus, Caporaso, and Knight 2005), a lack of coherence that is hard to reconcile with these interactions being the physical basis of the code.

\section{The adaptive theory: evidence of evolutionary optimization of the code}

Quantitative evidence in support of the translation-error minimization hypothesis has been inferred from comparison of the standard code with random alternative codes. For any code its cost can be calculated using the following formula:
\begin{equation}
\varphi(a(c))=\sum_c\sum_{c'}p(c'|c)d(a(c'),a(c)),
\end{equation}
where $a(c):A\to C$ is a given code, i.e., mapping of 64 codons $c\in C$ to 20 amino acids and stop signal $a(c)\in A$; $p(c'|c)$  is the relative probability to misread codon $c$  as codon $c'$, and $d(a(c'),a(c))$ is the cost associated with the exchange of the cognate amino acid $a(c)$  with the misincorporated amino acid $a(c')$. Under this approach, the less the cost $\varphi(a(c))$  the more robust the code is with respect to mistranslations, i.e., the greater the code's fitness.

The first reasonably reliable numerical estimates of the fraction of random codes that are more robust than the standard code have been obtained by Haig and Hurst (Haig and Hurst 1991) who showed that, under the assumption that any misreadings between two codons that differ by one nucleotide are equally probable, and if the polar requirement scale (Woese et al. 1966b) is employed as the measure of physicochemical similarity of amino acids, the probability of a random code to be fitter than the standard one is $P_1\approx 10^{-4}$. Using a refined cost function that took into account the non-uniformity of codon positions and base-dependent transition bias, Freeland and Hurst have shown that the fraction of random codes that outperforms the standard one is $P_2\approx 10^{-6}$, i.e., ``the genetic code is one in a million'' (Freeland 1998). Subsequent analyses have yielded even higher estimates of error minimization of the standard code (Freeland et al. 2000; Gilis et al. 2001; Freeland, Wu, and Keulmann 2003; Goodarzi, Nejad, and Torabi 2004).

Despite the convincing demonstration of the high robustness to misreadings of the standard code, the translation-error minimization hypothesis seems to have some inherent problems. First, to obtain any estimate of a code's robustness, it is necessary to specify the exact form of the cost function (1) that, even in its simplest form, consists of a specific matrix of codon misreading probabilities and specific costs associated with the amino acid substitutions. The form of the matrix $p(c'|c)$ proposed by Freeland et al. (Freeland 1998) is widely used (e.g., (Gilis et al. 2001; Zhu, Zeng, and Huang 2003; Archetti 2004; Goodarzi, Nejad, and Torabi 2004; Novozhilov, Wolf, and Koonin 2007)) but the supporting data are scarce. In particular, it has been convincingly shown that mistranslation in the first and third codon positions is more common than in the second position (Woese 1965b; Parker 1989; Kramer and Farabaugh 2007), but the transitional biased misreading in the second position is hard to justify from the available data. In part, to overcome this problem, Ardell and Sella formulated the first population-genetic model of code evolution where the changes in genomic content of a population are modeled along with the code changes (Ardell 1998; Ardell and Sella 2002; Sella and Ardell 2006). This approach is a generalization of the adaptive concept of code evolution that unifies the lethal-mutation and translation-error minimization hypotheses and incorporates the well-known fact that, among mutations, transitions are far more frequent than transversions (Collins and Jukes 1994; Kumar 1996). Essentially, the Ardell-Sella model describes coevolution of a code with genes that utilize it to produce proteins and explicitly takes into account the ``freezing effect'' of genes on a code that is due to the massive deleterious effect of code changes (Ardell and Sella 2002). Under this model, evolving codes tend to ``freeze'' in structures similar to that of the standard code and having similar levels of robustness.   

Another problem with the function (1) is that it relies on a measure of physicochemical similarity of amino acids. It is clear that any one such measure cannot be totally adequate. The amino acid substitution matrices such as PAM that are commonly used for amino acid sequence comparison appear not to be suitable for the study of the code evolution because these matrices have been derived from comparison of protein sequences that are encoded by the standard code, and hence cannot be independent of that code (Di Giulio 2001b). Therefore one must use a code-independent matrix derived from a first-principle comparison of physic-chemical properties of amino acids, such as the polar requirement scale (Woese et al. 1966b). However, the number of possible matrices of this kind is enormous, and there are no clear criteria for choosing the ``best'' one. Thus, arbitrariness is inherent in the matrix selection, and its effect on the conclusions on the level of optimization of a code is hard to assess.  

A potentially serious objection to the error-minimization hypothesis (Di Giulio 2000) is that, although the estimates of $P_1$ and $P_2$ indicate that the standard code outperforms most random alternatives, the number of possible codes that are fitter (more robust) than the standard one is still huge (it should be noted that estimates of the code robustness rely on  the employed randomization procedure; the one most frequently used involves shuffling of amino acid assignments between the synonymous codon series that are intrinsic to the standard code, so that $20!\approx 2.4\times 10^{18}$ possible codes are searched; different random code generators can produce substantially different results (Novozhilov, Wolf, and Koonin 2007)). It has been suggested that, if selection for minimization of translation error effect was the principal force of code evolution, the relative optimization level for the standard code would be significantly higher than observed (Di Giulio, Capobianco, and Medugno 1994). The counter argument offered by supporters of the error-minimization hypothesis is that the distribution of random code costs is bell-shaped, where more robust codes form a long tail, so because the process of adaptation is non-linear, approaching the absolute minimum is highly improbable (Freeland, Wu, and Keulmann 2003).

It has been suggested that the apparent code robustness could be a by-product of evolution that was driven by selective forces that have nothing to do with error minimization (Stoltzfus and Yampolsky 2007).  Specifically, it has been shown that the non-random assignments of amino acids in the standard code can be almost completely explained by incremental code evolution by codon capture or ambiguity reduction processes. However, this conclusion relies on the exact order of amino acids recruitment to the genetic code (Trifonov 2000; Trifonov 2004), primarily, on a specific interpretation of the evolution of biosynthetic pathways for amino acids, which remains a controversial issue. 

\section{What is the level of code optimization and how could the code get there?}

Regardless of the exact nature of the selective forces that had the greatest effect on the evolution of the code, it is a fact that the standard code is substantially robust to translational misreadings as well as mutations. Thus, is seems to be of considerable importance to determine, as objectively as possible, the level of the code's optimization. Intriguing questions associated with this problem are how much evolution the standard code underwent and what would be the most likely starting point for such evolution.
 
\begin{figure}[tb!]
\centering
\includegraphics[width=0.6\textwidth]{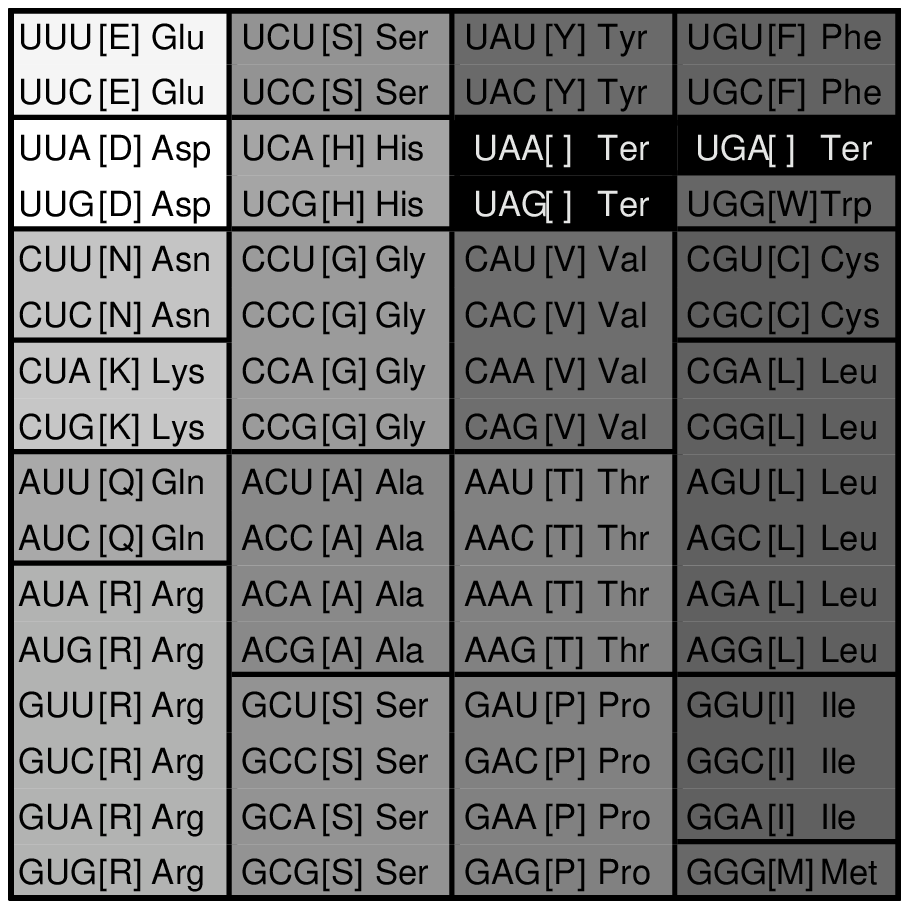}
\caption{An optimized genetic code with the same block structure and degeneracy as the standard code obtained as a result of combinatorial optimization of the amino acid assignments to four- and two-codon series. The optimization was performed by using the Great Deluge algorithm (Dueck 1993). The codon series are shaded in accordance with the polar requirement scale values as in Fig. 1}
\end{figure}

Estimates on the total level of code optimization have a long history. The straightforward comparison can be made between the standard code and the most robust code with respect to the mean cost value of random codes. This measure of the optimization level was dubbed the minimization percentage (Wong 1980; Di Giulio 1989); more precisely, $$MP=(\varphi_{mean}-\varphi_{stand})/(\varphi_{mean}-\varphi_{min}),$$ where $\varphi_{mean}$ is the mean cost of random codes, $\varphi_{stand}$ is the cost of the standard code, $\varphi_{min}$ is the cost of the most optimal code [all values are calculated given a particular cost function of the form (1)]. The minimization percentage of the standard code has been estimated at $\approx70\%$ when the polar requirement scale is used as the measure of amino acid exchangeability (Di Giulio 1989; Di Giulio, Capobianco, and Medugno 1994). Fig. 2 shows an example of a code that was optimized for robustness to translation errors by swapping codon assignments for amino acids to minimize the value of the cost function given by formula (1).  With respect to this code, the minimization percentage of the standard code is 78\% (this $MP$ value is somewhat higher than those reported by Di Giulio (Di Giulio, Capobianco, and Medugno 1994) because a more realistic misreading matrix $p(c'|c)$ was employed).

\begin{figure}[t!]
\centering
\includegraphics[width=0.7\textwidth]{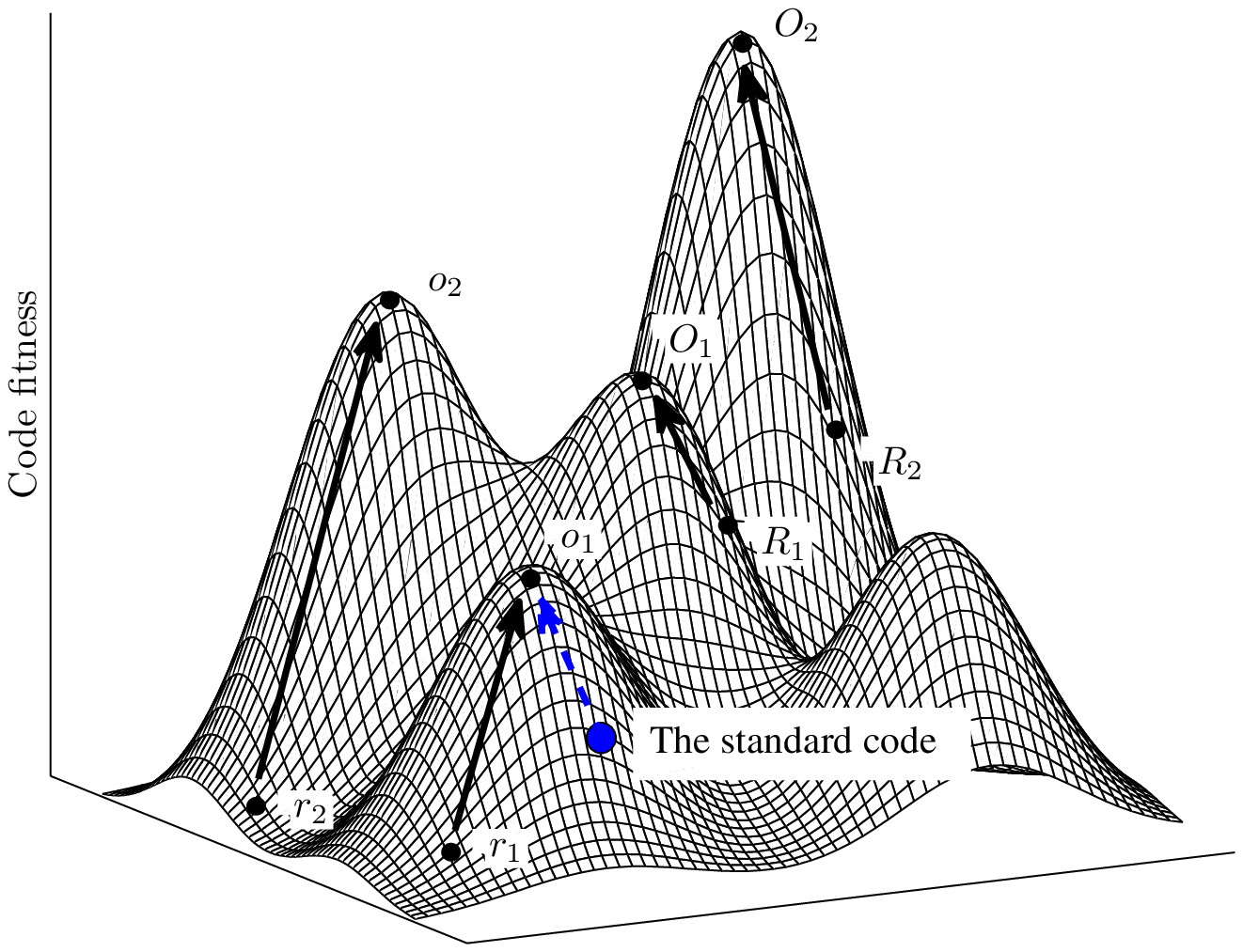}
\caption{Evolution of codes in a rugged fitness landscape (a cartoon illustration). $r_1,r_2\in \textbf{r}$ -- random codes with the same block structure as the standard code;  $o_1,o_2\in \textbf{o}$ -- codes obtained from  $\textbf{r}$  after optimization;  $R_1,R_2\in \textbf{R}$ -- random codes with fitness values greater than the fitness of the standard code;  $O_1,O_2\in \textbf{O}$ -- codes obtained from $\textbf{R}$ after optimization. The figure is modified from (Novozhilov, Wolf, and Koonin 2007)}
\end{figure}

Recently, we explored possible evolutionary trajectories of the genetic code within a limited domain of the vast space of possible codes (only codes that possess the same block structure and the same level of degeneracy as the standard code were analyzed) (Novozhilov, Wolf, and Koonin 2007). The assumption behind the choice of this small part of the vast code space is that, at an early stage of the evolution of the code, its block structure was fixed (``froze'') in the current form that could not be changed without a dramatic deleterious effect (a notion that is obviously related to Crick's frozen accident). Thus, we employed a straightforward, greedy evolutionary algorithm, with elementary steps comprising swaps of amino acid assignments between four-codon or two-codon series, to investigate the level of code optimization. The properties of the standard code were compared with the properties of four sets of random codes (purely random codes, random codes whose robustness is greater than that of the standard code, and two sets of codes that resulted from optimization of the first two sets). Under this model, the code fitness landscape is extremely rugged, so that almost any random code yields its own local maximum. Rather unexpectedly, starting from a random code, the level of optimization of the standard code can be easily achieved with 10-12 evolutionary steps on average, and often, optimization can be continued to reach the level that is attainable when the optimization starts from the standard code. When the starting point is a random code that is more robust than the standard one, the optimization procedure yields much higher levels of optimization than that reachable from the standard code, i.e., the standard code is much closer to its local fitness peak than most of the random codes with similar levels of robustness. Comparison of the standard code with the four described sets of codes shows that the standard code is very close to the set of optimized random codes. Thus, the standard genetic code appears to be a point that is located about half way (measured in the number of codon series swaps) along an upward evolutionary trajectory from a random code to the summit of the respective local peak. Moreover, this peak is rather mediocre, with a huge number of taller peaks existing in the landscape (Fig. 3). It should be emphasized that, under this model, the standard code is not locally stable, that is, it can be readily ``improved'' by a small perturbation (an additional swap). Thus, under the assumption that the function (1) is an adequate measure of the code fitness, it is hard to attribute the lack of further optimization of the standard code to anything other than frozen accident. 

\section{Coevolution theory: a link between the code and amino acid metabolism?}

The coevolution theory (reviewed in (Di Giulio 2004; Wong 2005; Wong 2007)) postulates that prebiotic synthesis could not produce 20 modern amino acids, so a subset of the amino acids had to be produced through biosynthetic pathways before they could be co-opted into the genetic code and translation; hence coevolution of the code and amino acid metabolism (Wong and Bronskill 1979). Therefore codon allocations to amino acids could have been guided by metabolic connections between the amino acids. According to the coevolution theory, there were three main phases of amino acid entry into the genetic code: the first (phase 1) amino acids came from prebiotic synthesis, phase 2 amino acids entered the code by means of biosynthesis from the phase 1 amino acids, and phase 3 amino acids are introduced into proteins through post-translational modifications (Wong 1981). The particular choice of phase 1 amino acids (Fig. 4) is supported by a survey of a variety of criteria used to infer the likely order of amino acid appearance (Trifonov 2000) (with one exception), and by the list of amino acids produced by high energy proton irradiation of a carbon monoxide-nitrogen-water mixture (Kobayashi et al. 1990). Under the coevolution theory, evolution of metabolic pathways is an important source of new amino acids. Given the precursor-product pairs of amino acids, the allocation of amino acids in the standard code is almost impossible to obtain by chance (Fig. 4). Experiments demonstrating that the amino acid composition of proteins is evolvable are construed as supporting the coevolution theory. For instance, it has been shown that \textit{Bacillus subtilis} could be mutated to replace its tryptophan by 4-fluoroTrp, and even further to displace Trp completely (Wong 1983).
\begin{figure}[bth]
\centering
\includegraphics[width=0.9\textwidth]{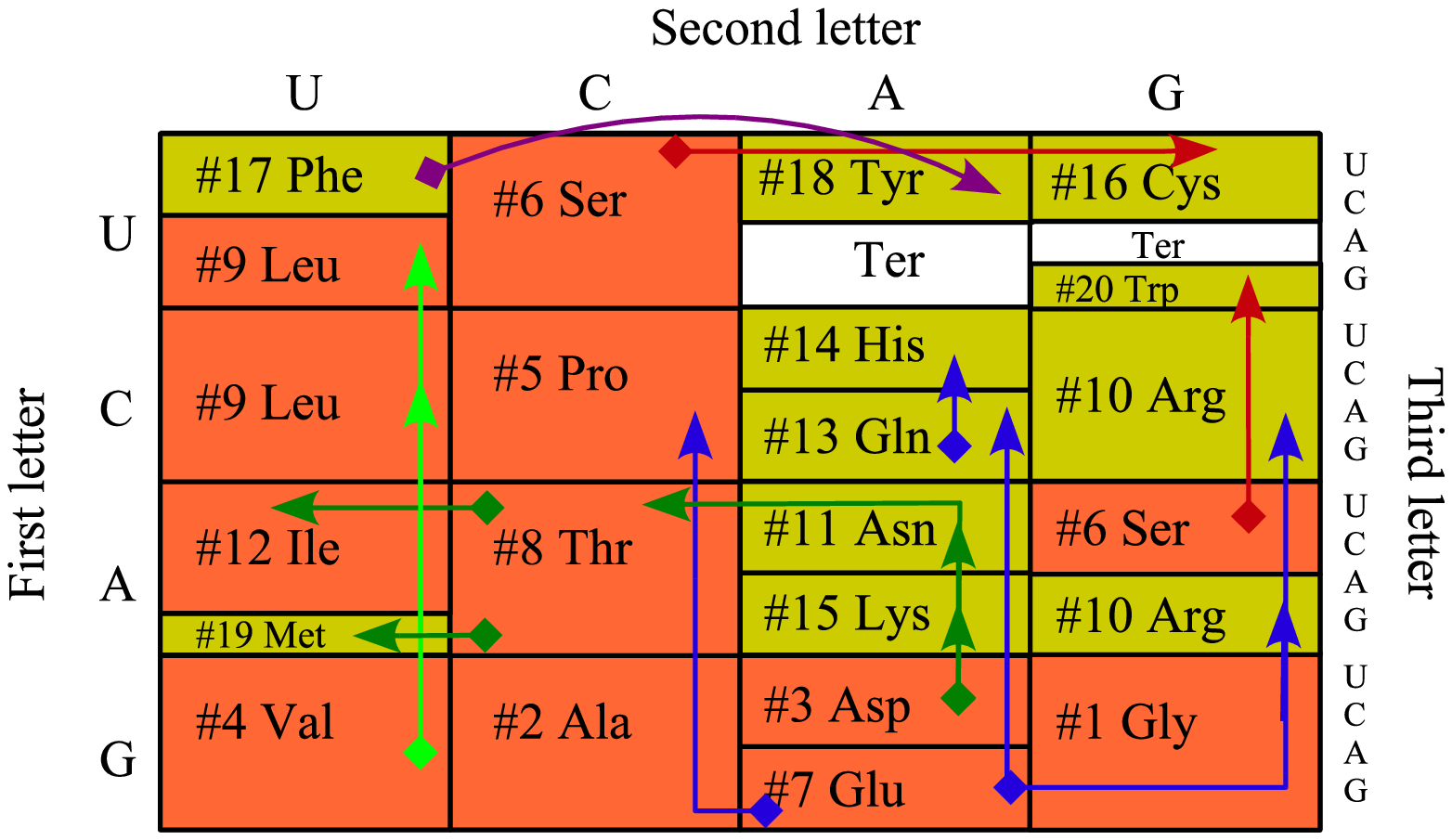}
\caption{The expansion of the standard code according to the coevolution theory. Phase 1 amino acids are orange, and phase 2 amino acids are green. The numbers show the order of amino acid appearance in the code according to (Trifonov 2004). The arrows define 13 precursor-product pairs of amino acids, their color defines the biosynthetic families of Glu (blue), Asp (dark-green), Phe (magenta), Ser (red), and Val (light-green)}
\end{figure}

Two major criticisms of the coevolution theory have been put forward. First, the coevolution scenario is very sensitive to the choice of amino acid precursor-product pairs, and the choice of these pairs is far from being straightforward. Indeed, in the original formulation of  the coevolution theory, Wong  did not directly use biochemically established relationships between amino acids but instead employed inferred reactions of primordial metabolism that remain debatable (Wong 1975; Wong 2007). Amirnovin (Amirnovin 1997) generated a large set of random codes and found that, if the original 8 precursor-product pairs proposed by Wong (Wong 1975) are considered, the standard code shows a substantially higher codon correlation score (a measure that calculates number of adjacent codons coding for precursor-product amino acids) than most of the random codes (only 0.1\% of random codes perform better). However, after the pairs Gln-His and Val-Leu are removed (the validity of the latter pair has been questioned (Ronneberg, Landweber, and Freeland 2000)), the proportion of better random codes rises to 3.6\%, and if the precursor-product pairs are taken from the well-characterized metabolic pathways of \textit{E. coli}, the proportion that a random code shows a stronger correlation reaches 34\%. Second, the biological validity of the statistical analysis of Wong (Wong 1975) appears dubious (Ronneberg, Landweber, and Freeland 2000). Ronneberg et al., together with consistent definition of amino acid precursor-product pairs, suggested that, according to the wobble rule, the genetic code contains not 61 functional codons coding for amino acids, but 45 codons, where each two codons of the form NNY are considered as one because no known tRNA can distinguish codons with U or C in the third base position. Under this assumption, no statistical support for the coevolution scenario of the evolution of the code was found (Ronneberg, Landweber, and Freeland 2000) (but see (Di Giulio 2001a)).  

\section{Is a compromise scenario plausible?}
As discussed above, despite a long history of research and accumulation of considerable circumstantial evidence, none of the three major theories on the nature and evolution of the genetic code is unequivocally supported by the currently available data. It appears premature to claim, e.g., that ``the coevolution theory is a proven theory'' (Wong 2007), or ``There is very significant evidence that cognate codons and/or anticodons are unexpectedly frequent in RNA-binding sites $[\ldots]$ This suggests that a substantial fraction of the genetic code has a stereochemical basis'' (Yarus, Caporaso, and Knight 2005). Is it conceivable that each of these theories captures some aspects of the code's origin and evolution, and combined, they could yield a more realistic picture?  In principle, it is not difficult to speculate along these lines, for instance, by imagining a scenario whereby first abiogenically synthesized amino acids captured their cognate codons owing to their respective stereochemical affinities, after which the code expanded according to the coevolution theory, and finally, amino acid assignments were adjusted under selection to minimize the effect of translational misreadings and point mutations on the genome. Such a composite theory is extremely flexible and consequently can ``explain'' just about anything by optimizing the relative contributions of different processes to fit the structure of the standard code. Of course, the falsifiability or, more generally, testability of such an overadjusted scenario become issues of concern. Nevertheless, examination of the specific predictions of each theory might take one some way toward falsification of the composite scenario.

The coevolution scenario implies that the genetic code should be highly robust to mistranslations, simply, because the identified precursor-product pairs consist of physico-chemically similar amino acids (Stoltzfus and Yampolsky 2007). However, several detailed analyses have suggested that coevolution alone cannot explain the observed level of robustness of the standard code so that additional evolution under selection for error minimization would be necessary to arrive to the standard code (Freeland and Hurst 1998; Freeland et al. 2000; Archetti 2004). Thus, in terms of the plausibility of a composite scenario, coevolution and error minimization are compatible. However, error minimization also appears to be necessary whereas the necessity of coevolution remains uncertain.

The affinities between cognate triplets and amino acids detected in aptamer selection experiments appear to be independent of the highly optimized amino acid assignments in the standard code table (Caporaso, Yarus, and Knight 2005). Thus, even if these affinities are relevant for the origin of the code, the error minimization properties of the standard code are still in need of an explanation. The proponents of the stereochemical theory argue that some of the amino acid assignments are stereochemically defined, whereas others have evolved under selective pressure for error minimization, resulting in the observed robustness of the standard code. Indeed, it has been shown that, even when 8-10 amino acid assignments in the standard code table are fixed, there is still plenty of room to produce highly optimized genetic codes (Caporaso, Yarus, and Knight 2005). However, this mixed stereochemistry--selection scenario seems to clash with some evidence. Perhaps, rather paradoxically, amino acids for which affinities with cognate triplets have been reported, largely, are considered to be late additions to the code: only 4 of the 8 amino acids with reported stereochemical affinities are phase 1 amino acids according to the coevolution theory (Fig. 4).  Notably, arginine, the amino acid for which the evidence in support of a stereochemical association with cognate codons appears to be the strongest, is the ``worst positioned'' amino acid in the code table, i.e., of all amino acids, a change in the codon  assignment for arginine results in the greatest increase in the code's fitness, e.g., (Novozhilov, Wolf, and Koonin 2007). This unusual position of arginine in the code table makes it tempting to consider a different combined scenario of the code's evolution whereby the early stage of this evolution involved, primarily, selection for error minimization, whereas at a later stage, the code was modified through recruitment of new amino acids that involved the (weak) stereochemical  affinities.

\section{Universality of the genetic code and collective evolution} 

Whether the code reflects biosynthetic pathways according to the coevolution theory or was shaped by adaptive evolutionary forces to minimize the burden caused by improper translated proteins or even to maximize the rate of the adaptive evolution of proteins (Maeshiro and Kimura 1998; Judson 1999; Zhu and Freeland 2005), a fundamental but often overlooked question is why the code is (almost) universal. Of course, the stereochemical theory, in principle, could offer a simple solution, namely, that the codon assignments in the standard code are unequivocally dictated by the specific affinity between amino acids and their cognate codons. As noticed above, however, the affinities are equivocal and weak, and do not account for the error-minimization property of the code.  An alternative could be that the code evolved to (near) perfection in terms of robustness to translational errors or, perhaps, some other optimization criteria, and this (nearly) perfect standard code outcompeted all other versions. We have seen, however, that, at least with respect to error minimization, this is far from being the case (Fig. 3). What remains as an explanation of the code's universality is some version of frozen accident combined with selection that brought the code to a relatively high robustness that was sufficient for the evolution of complex life. 

Under the frozen accident view, the universality of the code can be considered an epiphenomenon of the existence of a unique LUCA. The LUCA must have had a code with at least a minimal fitness compatible with cellular life, and that code was frozen ever since (except for the observed limited variation). The implicit assumption behind this line of reasoning is that LUCA already possessed a translation system that was (nearly) as advanced as the modern version. Indeed, the universality of the key components of the translation system including a nearly complete set of aminoacyl-tRNA synthetases among the extant cellular life forms (Harris et al. 2003; Koonin 2003) strongly suggests that the main features of the translation system were fixed at a pre-LUCA stage of evolution. 
The recently proposed hypothesis of collective evolution of primordial replicators explains the universality of the code through a combination of froze accident and a distinct type of selection pressure (Vetsigian, Woese, and Goldenfeld 2006; Goldenfeld and Woese 2007). The central idea is that universality of the genetic code is a condition for maintaining the (horizontal) flow of genetic information between communities of primordial replicators, and this information flow is a condition for the evolution of any complex biological entities. Horizontal transfer of replicators would provide the means for the emergence of clusters of similar codes, and these clusters would compete for niches. This idea of collective evolution of ensembles of virus-like genetic entities as a stage in the origin of cellular life apparently goes back to Haldane's classic paper of 1928 (Haldane 1928) but was subsequently recast in modern terms and expanded (Anderson 1970; Syvanen 1985; Woese 2000; Syvanen 2002), and developed in physical terms (Martin and Russell 2003; Koonin and Martin 2005). Vetsigian et al. (Vetsigian, Woese, and Goldenfeld 2006) explored the fate of the code under collective evolution using a simple evolutionary model which is a generalization of the population-genetic model of code evolution described by Sella and Ardell (Ardell and Sella 2002; Sella and Ardell 2006). It has been shown that, taking into consideration the selective advantage of error-minimizing codes, within a community of subpopulations of genetic elements capable of horizontal gene exchange, evolution leads to a nearly universal, highly robust code (Vetsigian, Woese, and Goldenfeld 2006). 

\section{Instead of conclusions: How did the code evolve (and will we ever know)?}

The writing of this review coincides with the $40^\textrm{th}$ anniversary of Crick's seminal paper on the evolution of the genetic code (Crick 1968) that synthesized the preceding research in this area and presciently outlined the principal lines of thinking on this difficult subject. In our opinion, despite extensive and, in many cases, elaborate attempts to model code optimization, ingenious theorizing along the lines of the coevolution theory, and considerable experimentation, very little definitive progress has been made.

Of course, this does not mean there has been no advance in understanding aspects of the code evolution. Some clear conclusions are negative, i.e., allow one to rule out certain \textit{a priori} plausible possibilities. Thus, many years of experimentation including the latest extensive studies on aptamer selection show that the code is not based on a straightforward stereochemical correspondence between amino acids and their cognate codons (or anticodons). Direct interactions between amino acids and polynucleotides might have been important at some early stages of code's evolution but hardly could have been the principal factor of the code's evolution. Almost the same seems to apply to the coevolution theory: the possibility exists that evolution of amino acid metabolism and evolution of the code were, to some extent, linked, but this coevolution cannot fully explain the properties of the code. The verdict on the adaptive theory of code evolution, in particular, the hypothesis that the code was shaped by selection for error minimization, is different: in our view, this is the only concept of the code evolution that can legitimately claim to be positively relevant as (so far) no attempt to explain the observed robustness of the code to translation errors without invoking at least some extent of selection has been convincing. So it does appear that selection for translation error minimization played a substantial role in the evolution of the code to the standard form. However, there is also a flip side to the adaptive theory as the standard code appears not to be particularly outstanding in terms of error minimization and, apparently, easily reachable from a random code with the same block structure. Statements like ``the genetic code is one in a million'' (or even in 100 million) are technically accurate but can be easily misconstrued should one overlook the fact that there is a huge number of possible codes that are significantly more robust than the standard code that sits on the slope of an unremarkable local peak in an extremely rugged fitness landscape (Fig. 3). Of course, it cannot be ruled out that the fitness functions employed in modeling selection for error minimization (eq. (1) and similar ones) in the evolution of the code are far from being an accurate representation of the ``real'' optimization criterion. Should that be the case, the general assessment of the entire field of code evolution would have to be particularly somber because that would imply we have no clue as to what is important in a code. However, this does not seem to be a particularly likely possibility. Indeed, recent theoretical and empirical studies on correlations between gene sequence evolution and expression strongly suggest that minimization of the production of potentially toxic misfolded proteins is a crucial factor of evolution (Drummond et al. 2005; Drummond, Raval, and Wilke 2006; Wilke and Drummond 2006; Drummond and Wilke 2008). It stands to reason that minimization of protein misfolding has driven evolution concordantly at several levels including protein sequences, codon usage (Drummond and Wilke 2008) and the genetic code itself. Furthermore, general considerations, stemming from Eigen's theory of quasispecies and mutational meltdown, indicate that, for any complex life to evolve, sufficient robustness of replication and expression is a pre-requisite (Zintzaras, Santos, and Szathmary 2002; Penny 2005; Wolf and Koonin 2007). Thus, these more general lines of reasoning from evolutionary biology seem to complement the results of specific modeling of the code's evolution.

And then, there is, of course, frozen accident, Crick's famous ``non-explanation'' that, even after 40 years of increasingly sophisticated research, still appears relevant for the problem of the code's origin and evolution. Indeed, given the relatively modest optimization level of the standard code, it appears essentially certain that the evolution of the code involved some combination of frozen accident with selection for error minimization. Whether or not other recognized and/or still unknown factors also contributed remains a matter to be addressed in further theoretical, modeling and experimental research.

Before closing this discussion, it makes sense to ask: do the analyses described here, focused on the properties and evolution of the code \textit{per se}, have the potential to actually solve the enigma of the code's origin? It appears that such potential is problematic because, out of necessity, to make the problems they address tractable, all studies of the code evolution are performed in formalized and, more or less, artificial settings (be it modeling under a defined set of code transformation or aptamer selection experiments) the relevance of which to the reality of primordial evolution is dubious at best. The hypothesis on the causal connection between the universality of the code and the collective character of primordial evolution characterized by extensive genetic exchange between ensembles of replicators (Vetsigian, Woese, and Goldenfeld 2006) is attractive and appears conceptually important because it takes the study of code evolution from being a purely formal exercise into a broader and more biologically meaningful context. Nevertheless, this proposal, even if quite plausible, is only one facet of a much more general and difficult problem, perhaps, the most formidable problem of all evolutionary biology. Indeed, it stands to reason that any scenario of the code origin and evolution will remain vacuous if not combined with understanding of the origin of the coding principle itself and the translation system that embodies it. At the heart of this problem is a dreary vicious circle: what would be the selective force behind the evolution of the extremely complex translation system before there were functional proteins? And, of course, there could be no proteins without a sufficiently effective translation system. A variety of hypotheses have been proposed in attempts to break the circle (see (Noller 2004; Penny 2005; Noller 2006; Wolf and Koonin 2007) and references therein) but so far none of these seems to be sufficiently coherent or enjoys sufficient support to claim the status of a real theory.

It seems that detailed modeling of the code evolution from simpler predecessors such as doublet codes could offer some new windows into the early stages of the evolution of coding (Delarue 2007). Notably, backtracking the standard code to the most likely doublet versions yields codes with an exceptional, nearly maximum error minimization capacity (ASN and EVK, unpublished), an observation that moves selection for error minimization and/or frozen accident at least one step closer to the actual origin of translation. Nevertheless, these and other theoretical approaches lack the ability to take the reconstruction of the evolutionary past beyond the complexity threshold that is required to yield functional proteins, and we must admit that concrete ways to cross that horizon are not currently known.

On the experimental front, findings on the catalytic capabilities of selected ribozymes are impressive (Fedor and Williamson 2005). In particular, highly efficient self-aminoacylating ribozymes and ribozymes that catalyze the peptidyltransferase reaction have been obtained (Illangasekare, Kovalchuke, and Yarus 1997; Cui, Sun, and Zhang 2004). Moreover, ribozymes whose catalytic activity is stimulated by peptides have been selected (Robertson, Knudsen, and Ellington 2004), hinting at the possible origins of the RNA-protein connection (Wolf and Koonin 2007). Nevertheless, in a close analogy to the situation with theoretical approaches, we are unaware of any experiments that would have the potential to actually reconstruct the origin of coding, not even at the stage of serious planning.

Summarizing the state of the art in the study of the code evolution, we cannot escape considerable skepticism. It seems that the two-pronged fundamental  question: ``Why is the genetic code the way it is and how did it come to be?'', that was asked over 50 years ago, at the dawn of molecular biology, might remain pertinent even in another 50 years. Our consolation is that we cannot think of a more fundamental problem in biology. 

\small{
\paragraph{Acknowledgements.} Although the study of the evolution of the genetic code is a relatively well focused field, the literature accumulated over the 50 years of research is extensive, and we could not possibly cover all of it in a brief review article. Our sincere apologies to all colleagues whose relevant work is not cited due to space restrictions. EVK is grateful to Nigel Goldenfeld, Paul Higgs, and Claus Wilke for insightful discussions during the workshop on ``Evolution: from Atoms to Organisms'' at the Aspen Center for Physics (Aspen, CO), 8/10/2008-8/31/2008. The authors' research is supported by the Department of Health and Human Services intramural program (NIH, National Library of Medicine).}\\[2mm]

\small 
\textbf{\large{References}}\\[1mm]

Aldana-Gonzalez, M., G. Cocho, H. Larralde, and G. Martinez-Mekler. 2003. Translocation Properties of Primitive Molecular Machines and Their Relevance to the Structure of the Genetic Code. {Journal of Theoretical Biology} 220:27-45.\\[-3mm]

Aldana, M., F. Cazarez-Bush, G. Cocho, and G. Martnez-Mekler. 1998. Primordial synthesis machines and the origin of the genetic code. Physica A 257:119-127.\\[-3mm]

Alff-Steinberger, C. 1969. The Genetic Code and Error Transmission. Proceedings of the National Academy of Sciences 64:584-591.\\[-3mm]

Alfonzo, J. D., V. Blanc, A. M. Estévez, M. A. T. Rubio, and L. Simpson. 1999. C to U editing of the anticodon of imported mitochondrial tRNA Trp allows decoding of the UGA stop codon in Leishmania tarentolae. The EMBO Journal 18:7056-7062.\\[-3mm]

Allmang, C., and A. Krol. 2006. Selenoprotein synthesis: UGA does not end the story. Biochimie 88:1561-1571.\\[-3mm]

Ambrogelly, A., S. Palioura, and D. Soll. 2007. Natural expansion of the genetic code. Nat Chem Biol 3:29-35.\\[-3mm]

Amirnovin, R. 1997. An Analysis of the Metabolic Theory of the Origin of the Genetic Code. Journal of Molecular Evolution 44:473-476.\\[-3mm]

Anderson, N. G. 1970. Evolutionary significance of virus infection. Nature 227:1346-1347.\\[-3mm]

Andersson, G. E., and C. G. Kurland. 1991. An extreme codon preference strategy: codon reassignment. Mol Biol Evol 8:530-544.\\[-3mm]

Andersson, S. G., and C. G. Kurland. 1995. Genomic evolution drives the evolution of the translation system. Biochem Cell Biol 73:775-787.\\[-3mm]

Andersson, S. G. E., and C. G. Kurland. 1998. Reductive evolution of resident genomes. Trends Microbiol 6:263-268.\\[-3mm]

Archetti, M. 2004. Codon Usage Bias and Mutation Constraints Reduce the Level of Error Minimization of the Genetic Code. Journal of Molecular Evolution 59:258-266.\\[-3mm]

Ardell, D. H. 1998. On error minimization in a sequential origin of the standard genetic code. J Mol Evol 47:1-13.\\[-3mm]

Ardell, D. H., and G. Sella. 2002. No accident: genetic codes freeze in error-correcting patterns of the standard genetic code. Philos Trans R Soc Lond B Biol Sci 357:1625-1642.\\[-3mm]

Barrell, B. G., A. T. Bankier, and J. Drouin. 1979. A different genetic code in human mitochondria. Nature 282:189-194.\\[-3mm]

Caporaso, J. G., M. Yarus, and R. Knight. 2005. Error Minimization and Coding Triplet/ Binding Site Associations Are Independent Features of the Canonical Genetic Code. Journal of Molecular Evolution 61:597-607.\\[-3mm]

Chechetkin, V. R. 2006. Genetic code from tRNA point of view. J Theor Biol 242:922-934.\\[-3mm]

Chechetkin, V. R. 2003. Block structure and stability of the genetic code. J Theor Biol 222:177-188.\\[-3mm]

Collins, D. W., and T. H. Jukes. 1994. Rates of transition and transversion in coding sequences since the human-rodent divergence. Genomics 20:386-396.\\[-3mm]

Crick, F. H. 1968. The origin of the genetic code. J Mol Biol 38:367-379.
Cui, Z., L. Sun, and B. Zhang. 2004. A peptidyl transferase ribozyme capable of combinatorial peptide synthesis. Bioorg Med Chem 12:927-933.\\[-3mm]

Davies, J., W. Gilbert, and L. Gorini. 1964. Streptomycin, Suppression, and the Code. Proceedings of the National Academy of Sciences 51:883-890.\\[-3mm]

Delarue, M. 2007. An asymmetric underlying rule in the assignment of codons: possible clue to a quick early evolution of the genetic code via successive binary choices. RNA 13:161-169.\\[-3mm]

Di Giulio, M. 1989. The extension reached by the minimization of the polarity distances during the evolution of the genetic code. Journal of Molecular Evolution 29:288-293.\\[-3mm]

Di Giulio, M. 2000. The origin of the genetic code. Trends in Biochemical Sciences 25:44.\\[-3mm]

Di Giulio, M. 2001a. A Blind Empiricism Against the Coevolution Theory of the Origin of the Genetic Code. Journal of Molecular Evolution 53:724-732.\\[-3mm]

Di Giulio, M. 2005. The origin of the genetic code: theories and their relationships, a review. Biosystems 80:175-184.\\[-3mm]

Di Giulio, M. 2001b. The Origin of the Genetic Code cannot be Studied using Measurements based on the PAM Matrix because this Matrix Reflects the Code Itself, Making any such Analyses Tautologous. Journal of Theoretical Biology 208:141-144.\\[-3mm]

Di Giulio, M. 2004. The coevolution theory of the origin of the genetic code. Physics of Life Reviews 1:128-137.\\[-3mm]

Di Giulio, M., M. R. Capobianco, and M. Medugno. 1994. On the optimization of the physicochemical distances between amino acids in the evolution of the genetic code. J Theor Biol 168:43-51.\\[-3mm]

Drummond, D. A., J. D. Bloom, C. Adami, C. O. Wilke, and F. H. Arnold. 2005. Why highly expressed proteins evolve slowly. Proc Natl Acad Sci U S A 102:14338-14343.\\[-3mm]

Drummond, D. A., A. Raval, and C. O. Wilke. 2006. A single determinant dominates the rate of yeast protein evolution. Mol Biol Evol 23:327-337.\\[-3mm]

Drummond, D. A., and C. O. Wilke. 2008. Mistranslation-induced protein misfolding as a dominant constraint on coding-sequence evolution. Cell 134:341-352.\\[-3mm]

Dueck, G. 1993. New optimization heuristics: the great deluge algorithm and the record-to-record travel. Journal of Computional Physics 104:86-92.\\[-3mm]

Dunnill, P. 1966. Triplet nucleotide-amino-acid pairing; a stereochemical basis for the division between protein and non-protein amino-acids. Nature 210:1267-1268.\\[-3mm]

Ellington, A. D., M. Khrapov, and C. A. Shaw. 2000. The scene of a frozen accident. RNA 6:485-498.\\[-3mm]

Epstein, C. J. 1966. Role of the amino-acid "code" and of selection for conformation in the evolution of proteins. Nature 210:25-28.\\[-3mm]

Fedor, M. J., and J. R. Williamson. 2005. The catalytic diversity of RNAs. Nat Rev Mol Cell Biol 6:399-412.\\[-3mm]

Freeland, S. J. 1998. The Genetic Code Is One in a Million. Journal of Molecular Evolution 47:238-248.\\[-3mm]

Freeland, S. J., and L. D. Hurst. 1998. Load minimization of the genetic code: history does not explain the pattern. Proceedings of the Royal Society B: Biological Sciences 265:2111-2119.\\[-3mm]

Freeland, S. J., R. D. Knight, L. F. Landweber, and L. D. Hurst. 2000. Early Fixation of an Optimal Genetic Code. Molecular Biology and Evolution 17:511-518.\\[-3mm]

Freeland, S. J., T. Wu, and N. Keulmann. 2003. The case for an error minimizing standard genetic code. Orig Life Evol Biosph 33:457-477.\\[-3mm]

Friedman, S. M., and I. B. Weinstein. 1964. Lack of fidelity in the translation of ribopolynucleotides. Proc. Natl. Acad. Sci. USA 52:988-996.\\[-3mm]

Gamow, G. 1954. Possible relation between deoxyribonucleic acid and protein structures. Nature 173:318.\\[-3mm]

Giege, R., M. Sissler, and C. Florentz. 1998. Universal rules and idiosyncratic features in tRNA identity. Nucleic Acids Research 26:5017-5035.\\[-3mm]

Gilis, D., S. Massar, N. J. Cerf, and M. Rooman. 2001. Optimality of the genetic code with respect to protein stability and amino-acid frequencies. Genome Biol 2:49.41-49.12.\\[-3mm]

Goldberg, A. L., and R. E. Wittes. 1966. Genetic Code: Aspects of Organization. Science 153:420.\\[-3mm]

Goldenfeld, N., and C. Woese. 2007. Connections Biology's next revolution. Nature 445:369.\\[-3mm]

Goodarzi, H., H. A. Nejad, and N. Torabi. 2004. On the optimality of the genetic code, with the consideration of termination codons. Biosystems 77:163-173.\\[-3mm]

Gusev, V. A., and D. Schulze-Makuch. 2004. Genetic code: Lucky chance or fundamental law of nature? Physics of Life Reviews 1:202-229.\\[-3mm]

Haig, D., and L. D. Hurst. 1991. A quantitative measure of error minimization in the genetic code. Journal of Molecular Evolution 33:412-417.\\[-3mm]

Haldane, J. B. S. 1928. The Origin of Life. Rationalist Annual 148:3-10.\\[-3mm]

Harris, J. K., S. T. Kelley, G. B. Spiegelman, and N. R. Pace. 2003. The genetic core of the universal ancestor. Genome Res 13:407-412.\\[-3mm]

Hasegawa, M., and T. Miyata. 1980. On the Asymmetry of the Amino Acid Code Table. Orig. Life 10:265-270.\\[-3mm]

Hinegardner, R. T., and J. Engelberg. 1963. Rationale for a Universal Genetic Code. Science 142:1083-1055.\\[-3mm]

Ikehara, K., and Y. Niihara. 2007. Origin and evolutionary process of the genetic code. Curr Med Chem 14:3221-3231.\\[-3mm]

Illangasekare, M., O. Kovalchuke, and M. Yarus. 1997. Essential structures of a self-aminoacylating RNA. J Mol Biol 274:519-529.\\[-3mm]

Itzkovitz, S., and U. Alon. 2007. The genetic code is nearly optimal for allowing additional information within protein-coding sequences. Genome Research 17:405-412.\\[-3mm]

Judson, O. P. 1999. The Genetic Code: What Is It Good For? An Analysis of the Effects of Selection Pressures on Genetic Codes. Journal of Molecular Evolution 49:539-550.\\[-3mm]

King, J. L., and T. H. Jukes. 1969. Non-Darwinian evolution. Science 164:788-798.\\[-3mm]

Knight, R. D., S. J. Freeland, and L. F. Landweber. 2001. Rewiring the keyboard: evolvability of the genetic code. Nat. Rev. Genet 2:49-58.\\[-3mm]

Knight, R. D., S. J. Freeland, and L. F. Landweber. 1999. Selection, history and chemistry: the three faces of the genetic code. Trends Biochem Sci 24:241-247.\\[-3mm]

Knight, R. D., and L. F. Landweber. 2000. Guilt by association: The arginine case revisited. RNA 6:499-510.\\[-3mm]

Knight, R. D., and L. F. Landweber. 1998. Rhyme or reason: RNA-arginine interactions and the genetic code. Chem. Biol 5:215-220.\\[-3mm]

Knight, R. D., L. F. Landweber, and M. Yarus. 2003. Tests of a Stereochemical Genetic Code. Pp. 115-128 in J. Lapointe, and L. Brakier-Gingras, eds. Translation Mechanism. Kluwer Academic/Plenum Publishers, New York.\\[-3mm]

Kobayashi, K., M. Tsuchiya, T. Oshima, and H. Yanagawa. 1990. Abiotic synthesis of amino acids and imidazole by proton irradiation of simulated primitive earth atmospheres. Origins of Life and Evolution of the Biosphere 20:99-109.\\[-3mm]

Koonin, E. V. 2003. Comparative genomics, minimal gene-sets and the last universal common ancestor. Nat Rev Microbiol 1:127-136.\\[-3mm]

Koonin, E. V., and W. Martin. 2005. On the origin of genomes and cells within inorganic compartments. Trends Genet 21:647-654.\\[-3mm]

Kramer, E. B., and P. J. Farabaugh. 2007. The frequency of translational misreading errors in E. coli is largely determined by tRNA competition. RNA 13:87-96.\\[-3mm]

Krzycki, J. A. 2005. The direct genetic encoding of pyrrolysine. Curr Opin Microbiol 8:706-712.\\[-3mm]

Kumar, S. 1996. Patterns of Nucleotide Substitution in Mitochondrial Protein Coding Genes of Vertebrates. Genetics 143:537-548.\\[-3mm]

Lu, Y., and S. Freeland. 2006. On the evolution of the standard amino-acid alphabet. Genome Biol 7:102.\\[-3mm]

Lu, Y., and S. J. Freeland. 2008. A quantitative investigation of the chemical space surrounding amino acid alphabet formation. J Theor Biol 250:349-361.\\[-3mm]

Maeshiro, T., and M. Kimura. 1998. The role of robustness and changeability on the origin and evolution of genetic codes. Proceedings of the National Academy of Sciences 95:5088-5093.\\[-3mm]

Martin, W., and M. J. Russell. 2003. On the origins of cells: a hypothesis for the evolutionary transitions from abiotic geochemistry to chemoautotrophic prokaryotes, and from prokaryotes to nucleated cells. Philos Trans R Soc Lond B Biol Sci 358:59-83.\\[-3mm]

Massey, S. E., and J. R. Garey. 2007. A Comparative Genomics Analysis of Codon Reassignments Reveals a Link with Mitochondrial Proteome Size and a Mechanism of Genetic Code Change Via Suppressor tRNAs. Journal of Molecular Evolution 64:399-410.\\[-3mm]

Massey, S. E., G. Moura, P. Beltrao, R. Almeida, J. R. Garey, M. F. Tuite, and M. A. S. Santos. 2003. Comparative Evolutionary Genomics Unveils the Molecular Mechanism of Reassignment of the CTG Codon in Candida spp. Genome Research 13:544-557.\\[-3mm]

Matsuyama, S., T. Ueda, P. F. Crain, J. A. McCloskey, and K. Watanabe. 1998. A novel wobble rule found in starfish mitochondria. Presence of 7-methylguanosine at the anticodon wobble position expands decoding capability of tRNA. Journal of Biological Chemistry 273:3363-3368.\\[-3mm]

Munteanu, A., C. S. Attolini, S. Rasmussen, H. Ziock, and R. V. Sole. 2007. Generic Darwinian selection in catalytic protocell assemblies. Philos Trans R Soc Lond B Biol Sci 362:1847-1855.\\[-3mm]

Nirenberg, M. W., W. Jones, P. Leder, B. F. C. Clark, W. S. Sly, and S. Pestka. 1963. On the coding of genetic information. Cold Spring Harb Symp Quant Biol 28:549-557.\\[-3mm]

Noller, H. F. 2004. The driving force for molecular evolution of translation. RNA 10:1833-1837.\\[-3mm]

Noller, H. F. 2006. Evolution of ribosomes and translation from an RNA world in R. F. Gesteland, T. R. Cech, and J. F. Atkins, eds. The RNA World. Cold Spring Harbor laboratory press, Cold Spring Harbor, NY.\\[-3mm]

Novozhilov, A. S., Y. I. Wolf, and E. V. Koonin. 2007. Evolution of the genetic code:  partial optimization of a random code for robustness to translation error in a rugged fitness landscape. Biology Direct 2(24).\\[-3mm]

Osawa, S. 1995. Evolution of the genetic code. Oxford University Press.\\[-3mm]

Osawa, S., T. H. Jukes, K. Watanabe, and A. Muto. 1992. Recent evidence for evolution of the genetic code. Microbiology and Molecular Biology Reviews 56:229-264.\\[-3mm]

Parker, J. 1989. Errors and alternatives in reading the universal genetic code. Microbiology and Molecular Biology Reviews 53:273-298.\\[-3mm]

Patel, A. 2005. The triplet genetic code had a doublet predecessor. J Theor Biol 233:527-532.\\[-3mm]

Pelc, S. R. 1965. Correlation between coding-triplets and amino acids. Nature 207:597-599.\\[-3mm]

Pelc, S. R., and M. G. E. Welton. 1966. Stereochemical relationship between coding triplets and amino-acids. Nature 209:868-870.\\[-3mm]

Penny, D. 2005. An interpretative review of the origin of life research. Biology and Philosophy 20:633-671.\\[-3mm]

Robertson, M. P., S. M. Knudsen, and A. D. Ellington. 2004. In vitro selection of ribozymes dependent on peptides for activity. RNA 10:114-127.\\[-3mm]

Ronneberg, T. A., L. F. Landweber, and S. J. Freeland. 2000. Testing a biosynthetic theory of the genetic code: Fact or artifact? Proceedings of the National Academy of Sciences 97:13690-13695.\\[-3mm]

Rumer, I. B. 1966. On codon systematization in the genetic code. Dokl Akad Nauk SSSR 167:1393-1394.\\[-3mm]

Santos, M. A. S., C. Cheesman, V. Costa, P. Moradas-Ferreira, and M. F. Tuite. 1999. Selective advantages created by codon ambiguity allowed for the evolution of an alternative genetic code in Candida spp. Molecular Microbiology 31:937-947.\\[-3mm]

Santos, M. A. S., G. Moura, S. E. Massey, and M. F. Tuite. 2004. Driving change: the evolution of alternative genetic codes. Trends Genet 20:95-102.\\[-3mm]

Saxinger, C., C. Ponnamperuma, and C. Woese. 1971. Evidence for the interaction of nucleotides with immobilized amino-acids and its significance for the origin of the genetic code. Nat New Biol 234:172-174.\\[-3mm]

Schultz, D. W., and M. Yarus. 1994. Transfer RNA mutation and the malleability of the genetic code. J Mol Biol 235:1377-1380.\\[-3mm]

Schultz, D. W., and M. Yarus. 1996. On malleability in the genetic code. J Mol Evol 42:597-601.\\[-3mm]

Sella, G., and D. H. Ardell. 2006. The coevolution of genes and genetic codes: Crick's frozen accident revisited. J Mol Evol 63:297-313.\\[-3mm]

Sengupta, S., X. Yang, and P. G. Higgs. 2007. The Mechanisms of Codon Reassignments in Mitochondrial Genetic Codes. Journal of Molecular Evolution 64:662-688.\\[-3mm]

Sonneborn, T. M. 1965. Degeneracy of the genetic code: extent, nature, and genetic implications. Evolving genes and proteins:377-397.\\[-3mm]

Stoltzfus, A., and L. Y. Yampolsky. 2007. Amino Acid Exchangeability and the Adaptive Code Hypothesis. J Mol Evol 65:456-462.\\[-3mm]

Suzuki, T., T. Ueda, and K. Watanabe. 1997. The ``polysemous'' codon --- a codon with multiple amino acid assignment caused by dual specificity of tRNA identity. The EMBO Journal 16:1122-1134.\\[-3mm]

Syvanen, M. 1985. Cross-species gene transfer; implications for a new theory of evolution. J Theor Biol 112:333-343.\\[-3mm]

Syvanen, M. 2002. Recent emergence of the modern genetic code: a proposal. Trends Genetics 18:245-248.\\[-3mm]

Szathmary, E. 2003. Why are there four letters in the genetic alphabet? Nat Rev Genet 4:995-1001.\\[-3mm]

Szathmary, E. 1991. Four Letters in the Genetic Alphabet: A Frozen Evolutionary Optimum? Proceedings: Biological Sciences 245:91-99.\\[-3mm]

Travers, A. 2006. The evolution of the genetic code revisited. Orig Life Evol Biosph 36:549-555.\\[-3mm]

Trifonov, E. N. 2000. Consensus temporal order of amino acids and evolution of the triplet code. Gene 261:139-151.\\[-3mm]

Trifonov, E. N. 2004. The triplet code from first principles. J Biomol Struct Dyn 22:1-11.\\[-3mm]

Vetsigian, K., C. Woese, and N. Goldenfeld. 2006. Collective evolution and the genetic code. Proceedings of the National Academy of Sciences 103:10696-10701.\\[-3mm]

Vol'kenshtein, M. V., and I. B. Rumer. 1967. Systematics of codons. Biofizika 12:10-13.\\[-3mm]

Wang, L., J. Xie, and P. G. Schultz. 2006. Expanding the genetic code. Annu Rev Biophys Biomol Struct 35:225-249.\\[-3mm]

Weber, A. L., and S. L. Miller. 1981. Reasons for the occurrence of the twenty coded protein amino acids. Journal of Molecular Evolution 17:273-284.\\[-3mm]

Wetzel, R. 1995. Evolution of the aminoacyl-tRNA synthetases and the origin of the genetic code. Journal of Molecular Evolution 40:545-550.\\[-3mm]

Wilke, C. O., and D. A. Drummond. 2006. Population genetics of translational robustness. Genetics 173:473-481.\\[-3mm]

Woese, C. R. 1965a. Order in the Genetic Code. Proceedings of the National Academy of Sciences 54:71-75.\\[-3mm]

Woese, C. R. 1967. The Genetic Code: The Molecular Basis for Genetic Expression. Harper \& Row.\\[-3mm]

Woese, C. R. 1965b. On the evolution of the genetic code. Proc Natl Acad Sci USA 54:1546-1552.\\[-3mm]

Woese, C. R. 2000. Interpreting the universal phylogenetic tree. Proceedings of the National Academy of Sciences 97:8392-8396.\\[-3mm]

Woese, C. R., D. H. Dugre, S. A. Dugre, M. Kondo, and W. C. Saxinger. 1966a. On the fundamental nature and evolution of the genetic code. Cold Spring Harb Symp Quant Biol 31:723-736.\\[-3mm]

Woese, C. R., D. H. Dugre, W. C. Saxinger, and S. A. Dugre. 1966b. The Molecular Basis for the Genetic Code. Proceedings of the National Academy of Sciences 55:966-974.\\[-3mm]

Woese, C. R., and G. E. Fox. 1977. The concept of cellular evolution. J Mol Evol 10:1-6.\\[-3mm]

Woese, C. R., R. T. Hinegardner, and J. Engelberg. 1964. Universality in the Genetic Code. Science 144:1030-1031.\\[-3mm]

Wolf, Y. I., and E. V. Koonin. 2007. On the origin of the translation system and the genetic code in the RNA world by means of natural selection, exaptation, and subfunctionalization. Biology Direct 2.\\[-3mm]

Wong, J. T. F. 1981. Coevolution of genetic code and amino acid biosynthesis. Trends Biochem. Sci 6:33-35.\\[-3mm]

Wong, J. T. F. 2007. Question 6: Coevolution Theory of the Genetic Code: A Proven Theory. Origins of Life and Evolution of Biospheres 37:403-408.\\[-3mm]

Wong, J. T. F. 1983. Membership Mutation of the Genetic Code: Loss of Fitness by Tryptophan. Proceedings of the National Academy of Sciences 80:6303-6306.\\[-3mm]

Wong, J. T. F. 1980. Role of Minimization of Chemical Distances between Amino Acids in the Evolution of the Genetic Code. Proceedings of the National Academy of Sciences 77:1083-1086.\\[-3mm]

Wong, J. T. F. 2005. Coevolution theory of the genetic code at age thirty. BioEssays 27:416-425.\\[-3mm]

Wong, J. T. F. 1975. A Co-Evolution Theory of the Genetic Code. Proceedings of the National Academy of Sciences 72:1909-1912.\\[-3mm]

Wong, J. T. F., and P. M. Bronskill. 1979. Inadequacy of prebiotic synthesis as origin of proteinous amino acids. Journal of Molecular Evolution 13:115-125.\\[-3mm]

Wu, H. L., S. Bagby, and J. M. van den Elsen. 2005. Evolution of the genetic triplet code via two types of doublet codons. J Mol Evol 61:54-64.\\[-3mm]

Xie, J., and P. G. Schultz. 2006. A chemical toolkit for proteins - an expanded genetic code. Nat Rev Mol Cell Biol 7:775-782.\\[-3mm]

Yarus, M., J. G. Caporaso, and R. Knight. 2005. Origins of the Genetic Code: The Escaped Triplet Theory. Annual Review of Biochemistry 74:179-198.\\[-3mm]

Ycas, M. 1969. The Biological Code. North-Holland, Amsterdam.\\[-3mm]

Yokobori, S., T. Suzuki, and K. Watanabe. 2001. Genetic Code Variations in Mitochondria: tRNA as a Major Determinant of Genetic Code Plasticity. Journal of Molecular Evolution 53:314-326.\\[-3mm]

Zhu, C. T., X. B. Zeng, and W. D. Huang. 2003. Codon Usage Decreases the Error Minimization Within the Genetic Code. Journal of Molecular Evolution 57:533-537.\\[-3mm]

Zhu, W., and S. Freeland. 2005. The standard genetic code enhances adaptive evolution of proteins. J Theor Biol 239:63-70.\\[-3mm]

Zintzaras, E., M. Santos, and E. Szathmary. 2002. "Living" under the challenge of information decay: the stochastic corrector model vs. hypercycles. J Theor Biol 217:167-181.

\end{document}